\documentclass[
    ,final            
  ]
  {aipproc}

\layoutstyle{6x9}

\begin{document}

\title{Highlights of the VERITAS Blazar Program}


\author{Wystan Benbow}{
  address={Harvard-Smithsonian Center for Astrophysics, 60 Garden St MS-20, Cambridge, MA 02138, USA}
}

\author{the VERITAS Collaboration}{
  address={\url{http://veritas.sao.arizona.edu/}}
}

\begin{abstract}

The VERITAS array of 12-m atmospheric-Cherenkov telescopes
in southern Arizona began full-scale operations in 2007,
and is one of the world's most-sensitive detectors of
astrophysical VHE (E>100 GeV) $\gamma$-rays.
Approximately 50 blazars are known to emit VHE photons, and observations of
blazars are a major focus of the VERITAS Collaboration.  
Nearly 2000 hours have been devoted to this program and $\sim$130 
blazars have already been observed with the array, 
in most cases with the deepest-ever
VHE exposure. These observations have resulted in
21 detections, including 10 VHE discoveries. 
Recent highlights of the VERITAS blazar observation program, 
and the collaboration's long-term blazar observation strategy, 
are presented.

\end{abstract}

\maketitle

\vspace{-0.2cm}
Blazars are a class of AGN with relativistic jets pointed along 
the line of sight to the observer, and are the most
numerous class of VHE source.  Both sub-classes of blazars, 
BL\,Lac objects and Flat Spectrum
Radio Quasars (FSRQs), are detected at VHE.  
Nearly 80\% of the VHE blazar population
exhibits VHE flux variability.  These variations
occur on time scales from minutes to years, but are typically
only a factor of 2 to 3.  The photon spectra of the
observed VHE emission is often soft ($\Gamma_{obs} \sim 2.5 - 4.6$), 
largely due to the attenuation of VHE 
photons on the extragalactic background light (EBL). 

Fermi-LAT has detected almost every VHE blazar.
By combining VHE and LAT spectra, the highest-energy peak of 
the double-humped blazar spectral energy distributions (SEDs)
is completely resolved.  By combining this
with X-ray and optical data, sampling the
lower-energy peak, it is possible to constrain source models.
VHE blazar studies can also lead to strong new constraints on the 
EBL and the intergalactic magnetic field (IGMF), and 
provide insight into the origin of ultra-high-energy cosmic rays.

Approximately 80\% of the known
VHE blazars are high-frequency-peaked BL\,Lac objects (HBLs).
The SEDs of these HBLs can usually be fit by a synchrotron 
self-Compton (SSC) model.  There are indications 
that the VHE emission of intermediate and low-frequency peaked 
BL Lac objects (IBLs and LBLs, respectively) has a different leptonic origin 
(SSC plus external Compton).  However, the current IBL/LBL detections 
are plagued by poor statistics at VHE,
and these non-HBL objects are typically only detected during flaring
episodes which might have a different physical origin.

\vspace{-0.5cm}
\section{The VERITAS Blazar Program}
\vspace{-0.3cm}

VERITAS \cite{Holder_ICRC} is sensitive to astrophysical $\gamma$-rays
in the energy range between $\sim$100 GeV and $\sim$30 TeV.  
The array is used to take $\sim$1000 h of good-weather data 
each year, and can detect a 1\% Crab Nebula flux 
source at zenith in $<$25 h.  
The typical systematic errors on VERITAS measurements 
are 0.1 for the photon index and 20\% in the integral flux.
An upgrade of VERITAS was completed
in 2012.  The new energy threshold is expected
to be $\sim$60 GeV with an overall sensitivity increase of $\sim$20\%.

The VERITAS collaboration hopes to improve the 
understanding of VHE blazars and their related science by
expanding the known population and by making precision
measurements of their spectra
and variability patterns.  Contemporaneous multi-wavelength (MWL) 
observations are a key component of this program since blazars
are highly-variable at all wavelengths.  
Modeling of the resulting SEDs and searches for MWL correlations
are used to provide insight into origin of the observed emission.

Since September 2007, VERITAS has observed
128 blazars, for a total of 1981 h (annual average of $\sim$400 h)
during good weather conditions.  Until 2010, $\sim$80\% of 
the VERITAS blazar data were used for VHE discovery 
observations and follow-up studies of new sources.  
In 2010, the focus became deep studies of known VHE sources, 
and $\sim$60\% of the data 
are now devoted to known sources.  Target-of-Opportunity
(ToO) observations are a key part of the blazar program and typically
average 50 h per year.  The 21 blazars detected by VERITAS are shown in 
Table~\ref{blazar_table}; see \cite{Benbow_ICRC11_disc,Benbow_ICRC11_KSP}
for more details.

\begin{table}[]
\begin{tabular}{c | c | c | c | c}
\hline
{\footnotesize Blazar} & {\footnotesize $z$} &  {\footnotesize Type} & 
{\footnotesize log$_{10}(\nu_{\rm synch})$ \cite{Nieppola}} 
& {\footnotesize Long-term Monitoring}\\
\hline
{\footnotesize Mrk\,421} & {\footnotesize 0.030} & {\footnotesize HBL} & 
{\footnotesize 18.5} & {\footnotesize Bright HBL (15 h) }\\
{\footnotesize Mrk\,501} & {\footnotesize 0.034} & {\footnotesize HBL} & 
{\footnotesize 16.8} & {\footnotesize Bright HBL (15 h)}\\
{\footnotesize 1ES\,2344+514} & {\footnotesize 0.044} & 
{\footnotesize HBL} & {\footnotesize 16.4} & {\footnotesize Bright HBL (5 h)}\\
{\footnotesize 1ES\,1959+650} & {\footnotesize 0.047} & {\footnotesize HBL} & 
{\footnotesize 18.0} & {\footnotesize Bright HBL (5 h)}\\
{\footnotesize BL\,Lac} & {\footnotesize 0.069} & {\footnotesize LBL} & 
{\footnotesize 14.3}& {\footnotesize Non-HBL (10 h)}\\
{\footnotesize W\,Comae$^{\dagger}$} & {\footnotesize 0.102} & 
{\footnotesize IBL} & {\footnotesize 14.8} & {\footnotesize Non-HBL (10 h)}\\
{\footnotesize RGB\,J0710+591$^{\dagger}$} & {\footnotesize 0.125} & 
{\footnotesize HBL} & {\footnotesize 21.1} & {\footnotesize EBL / IGMF  (25 h)}\\
{\footnotesize H\,1426+428} & {\footnotesize 0.129} & {\footnotesize HBL} & 
{\footnotesize 18.6} & {\footnotesize EBL / IGMF (25 h)}\\
{\footnotesize 1ES\,0806+524$^{\dagger}$} & {\footnotesize 0.138} & 
{\footnotesize HBL} & {\footnotesize 16.6} & {\footnotesize $-$ }\\
{\footnotesize 1ES\,0229+200} & {\footnotesize 0.140} & {\footnotesize HBL} & 
{\footnotesize 19.5} & {\footnotesize EBL / IGMF (25 h)}\\
{\footnotesize 1ES\,1440+122$^{\dagger}$} & {\footnotesize 0.162} & 
{\footnotesize IBL} & {\footnotesize 16.5} & {\footnotesize $-$ }\\
{\footnotesize RX\,J0648.7+1516$^{\dagger}$} & {\footnotesize 0.179} & 
{\footnotesize HBL} & {\footnotesize $-$ } & {\footnotesize $-$ }\\
{\footnotesize 1ES\,1218+304} & {\footnotesize 0.184} & {\footnotesize HBL} & 
{\footnotesize 19.1 } & {\footnotesize EBL / IGMF (25 h)}\\
{\footnotesize RBS\,0413$^{\dagger}$} & {\footnotesize 0.190} & 
{\footnotesize HBL} & {\footnotesize 17.0} & {\footnotesize $-$ }\\
{\footnotesize 1ES\,0414+009} & {\footnotesize 0.287} & {\footnotesize HBL} & 
{\footnotesize 20.7} & {\footnotesize EBL / IGMF (25 h)}\\
{\footnotesize PG\,1553+113} & {\footnotesize $0.43 < z < 0.50$} &
{\footnotesize HBL} & {\footnotesize 16.5} & {\footnotesize EBL / IGMF (25 h)}\\
{\footnotesize 3C\,66A$^{\dagger}$} & {\footnotesize ?} & {\footnotesize IBL} 
& {\footnotesize 15.6} & {\footnotesize Non-HBL (10 h)}\\
{\footnotesize B2\,1215+30} & {\footnotesize ?} & {\footnotesize IBL} & 
{\footnotesize 15.6} & {\footnotesize $-$ }\\
{\footnotesize PKS\,1424+240$^{\dagger}$} & {\footnotesize ?} & 
{\footnotesize IBL} & {\footnotesize 15.7} & {\footnotesize $-$ }\\
{\footnotesize 1ES\,0502+675$^{\dagger}$} & {\footnotesize ?} &
{\footnotesize HBL} & {\footnotesize 19.2} & {\footnotesize $-$ }\\
{\footnotesize RGB\,J0521.8+2112$^{\dagger}$} & {\footnotesize ?} & 
{\footnotesize HBL} & {\footnotesize $-$} & {\footnotesize $-$ }\\
\hline
\end{tabular}
\vspace{-0.2cm}
\caption{{\footnotesize The 21 blazars detected at VHE with VERITAS. The 10
VHE discoveries are marked with $\dagger$.  The long-term monitoring targets
are shown along with their reason for inclusion in the program and 
the current annual exposure goals. Both the targets and the exposures may evolve.
The LBL S5\,0716+714 is also monitored (10 h goal), but is not yet detected
by VERITAS.}}
\label{blazar_table}
\vspace{-0.3cm}
\end{table}

\vspace{-0.5cm}
\subsection{Recent Highlights from the VERITAS Blazar Program}
\vspace{-0.3cm}

\noindent VERITAS has regularly observed BL\,Lacertae (an LBL)
since the start of the 2010-11 season.  It is not usually 
bright enough to be detected at VHE.  However, in June 2011,
the tail of a bright VHE flare was observed.
The peak VHE flux observed was 125\% of the Crab Nebula flux, and the flare
exponentially decayed in $\tau = 13 \pm 4$ minutes.  
Despite being the brightest VHE flux ever observed from the object, 
the observed photon index ($\Gamma = 3.6 \pm 0.4$) is consistent 
with that measured by MAGIC during a lower-flux (3\% Crab Nebula)
flare in 2006. More details on the VERITAS and other MWL
observations of this exceptional flare can be found in these 
proceedings \cite{IBL_LBL_HDGS}.

\begin{figure*}[t]
   \centerline{
               \includegraphics[width=2.9in]{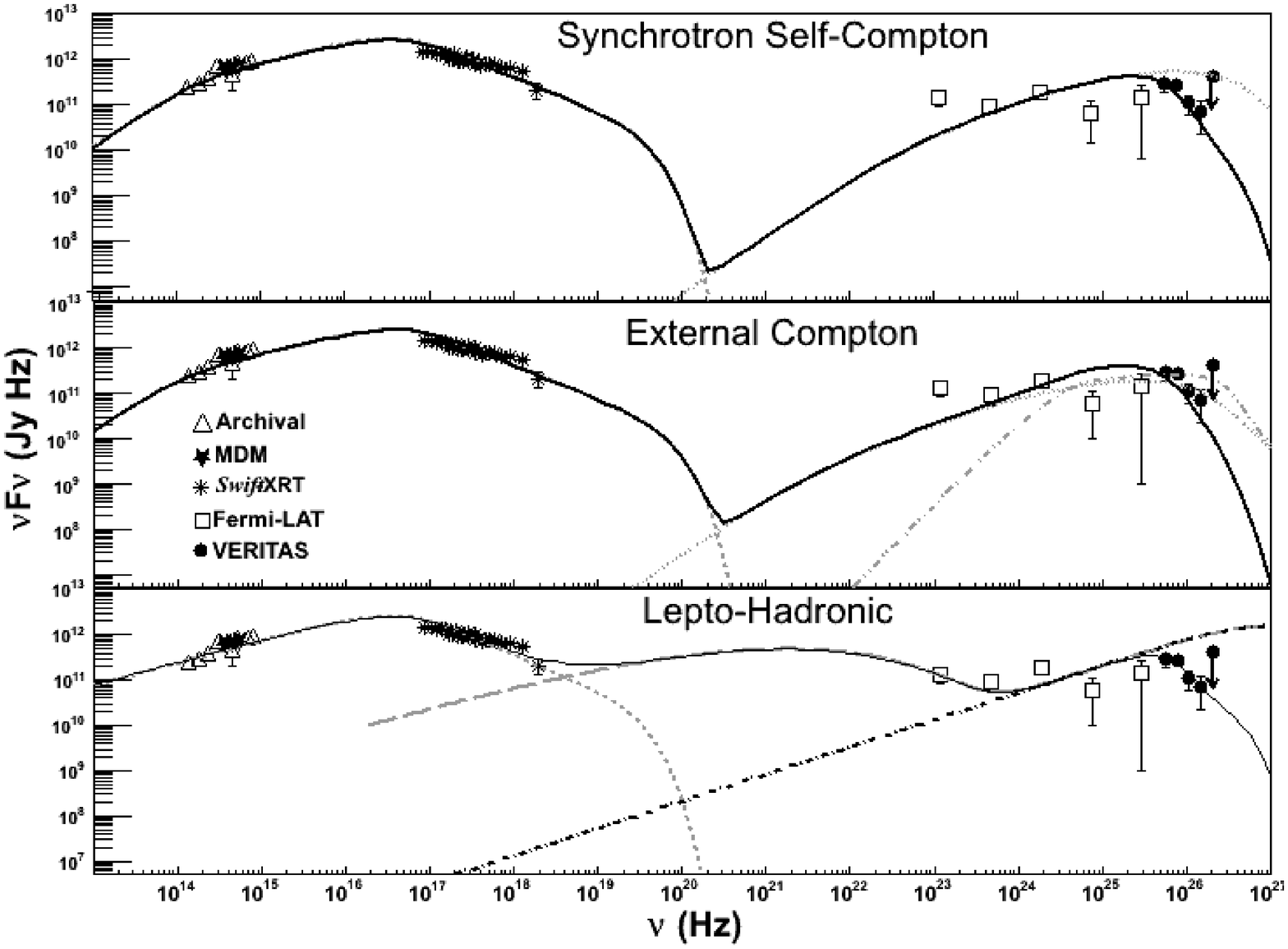}
		\hfil
               \includegraphics[width=3.0in]{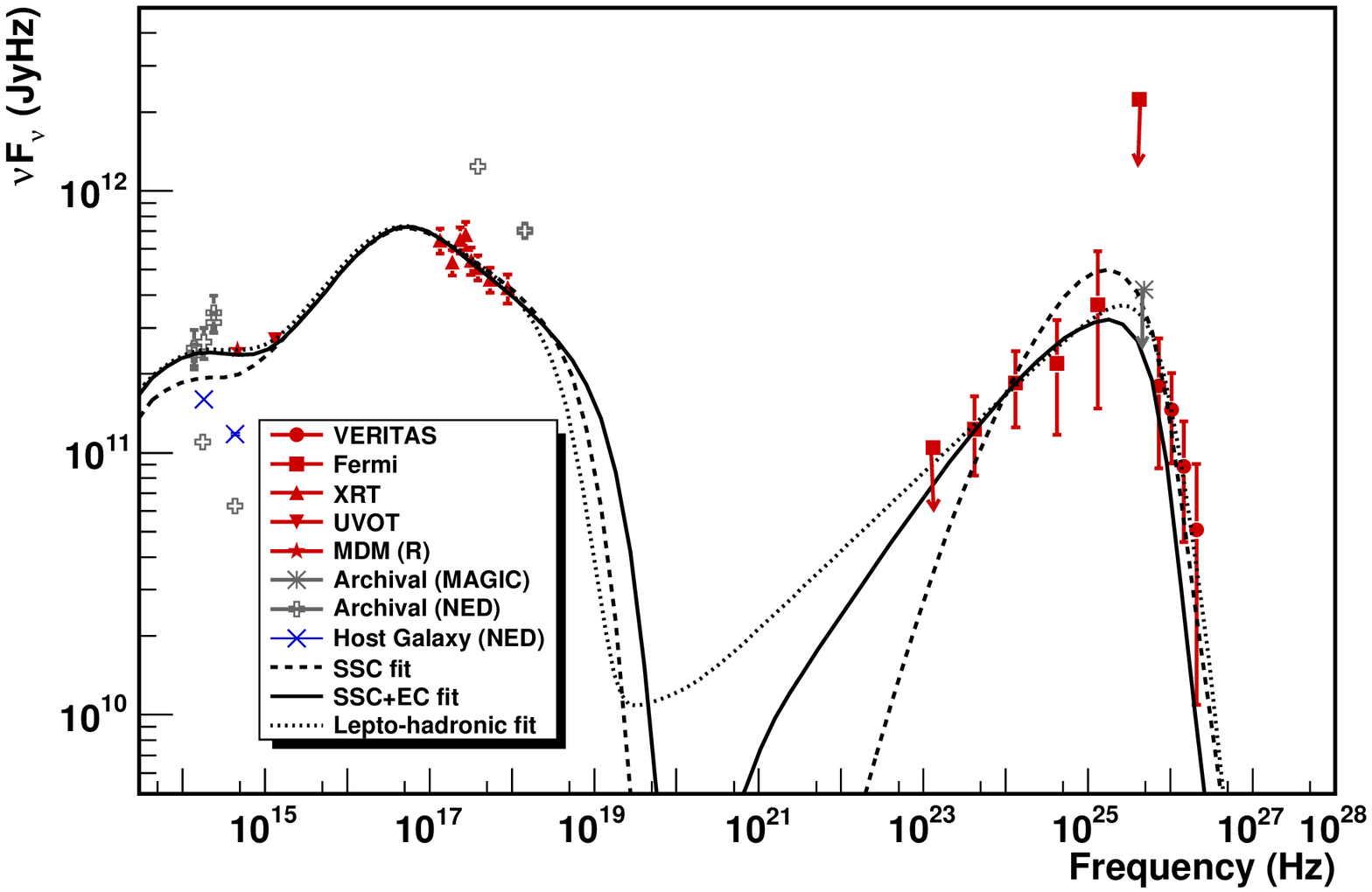}
              }
   \caption{\footnotesize The SEDs of (Left)
1ES\,0414+009 \cite{0414_paper}
and (Right) RBS\,0413 \cite{RBS0413_paper}.}
   \label{blazar_plots}
 \end{figure*}

The comparatively distant ($z = 0.287$) HBL 1ES\,0414+009 
is a known VHE emitter and was observed 
with VERITAS for 56 h of quality-selected live time between 2008
and 2011.  These data resulted in the significant detection of a source with flux $\sim$2\% 
of the Crab Nebula flux \cite{0414_paper}.  The observed spectrum is
soft $\Gamma = 3.4 \pm 0.5$, largely due to the effects of EBL
absorption.  Although an HBL, its SED (see Figure~\ref{blazar_plots})
is not well fit by one-zone leptonic models;
a lepto-hadronic model provides a better fit. 

Since September 2007, $\sim$1260 h of VERITAS data were 
devoted to the discovery and follow-up observations 
of more than 100 blazars at VHE.
The candidates observed are largely HBLs, but also include IBLs,
LBLs, FSRQs, and Fermi-LAT sources likely associated with blazars.
From 2007-09, the discovery program \cite{Benbow_ICRC09} focused on nearby
(redshift, $z < 0.3$) X-ray-bright HBLs, IBLs, and 
FSRQs recommended in the literature 
and nearby EGRET-detected blazars.  Some of the X-ray-bright targets 
were objects that met the selection criteria applied 
in the literature, but were first reported in later 
catalogs.
Following the release of the
first Fermi-LAT catalog of MeV-GeV-bright blazars in 2009 \cite{Fermi_LBAS},
the discovery program has focused on objects detected 
by Fermi-LAT \cite{2FGL_Catalog}.  

These observations resulted in the VERITAS discovery of 
VHE $\gamma$-ray emission from 6 HBLs
and the first 4 IBLs known to emit VHE photons \cite{Benbow_ICRC11_disc}. 
The initial observations of 60\% of these new VHE sources were motivated 
by the X-ray/EGRET-based selection criteria, and all
these objects are detected by Fermi-LAT.  In general, these
new sources have soft VHE spectra ($\Gamma_{avg} = 3.5$), and
only three exhibit strong VHE flux variability.  The
VHE flux upper limits derived from the unsuccesful discovery observations
are typically less than 2\% of the Crab Nebula 
flux and often the strictest yet.

The successful VHE discovery observations \cite{RXJ0648_paper} of 
RX\,J0648.7+1516 were motivated by the 
identification of a cluster of $E >10$ GeV photons in the Fermi-LAT data.
A total of 19 h of quality-selected live time was taken in March-April 2010.
Significant VHE emission ($\Gamma = 4.4 \pm 0.8$, $\sim$3\% Crab Nebula flux) was
detected at the position of RX\,J0648.7+1516, an unidentified radio, 
optical and X-ray emitter.  Follow-up optical spectroscopy
of this object yielded a continuum-dominated spectrum 
typical of BL Lac objects and weak absorption lines compatible with z = 0.179.
The SED of this blazar indicates it is an HBL, but it is not well fit by an SSC model.

RBS\,0413 is a bright HBL, at $z = 0.190$, that
was initially targeted based on its inclusion 
in the Sedentary Survey \cite{Sedentary}.
It was observed for $\sim$26 h of quality-selected live time between
2008 and 2010 \cite{RBS0413_paper}.  The data yielded the
detection of a soft-spectrum ($\Gamma = 3.2 \pm 0.7$) source with
$\sim$1\% Crab Nebula flux. 
Its SED, shown in Figure~\ref{blazar_plots}, is again
not well described by a SSC model.
The modeling favors the inclusion of an additional 
external-Compton (EC) component,
and can also be fit by a lepto-hadronic scenario. 

\vspace{-0.5cm}
\section{The Long-term Blazar Observation Strategy}
\vspace{-0.3cm}
In 2010, VERITAS began a systematic campaign to acquire 
deep exposures on 14 VHE blazars.   
Much of these data are acquired via regular monitoring exposures,
and data taking is intensified
during any flaring behavior.  Regular simultaneous 
MWL observations are also organized.  The 14 selected blazars are indicated
in Table 1, and include:
\vspace{-0.2cm}
\begin{itemize}
\item 6 comparatively distant, hard-VHE-spectrum HBLs for EBL and IGMF studies.

\item 4 bright HBLs to easily enable MWL studies with high VHE statistics. 
These may also have the highest likelihood
for exhibiting greater than Crab Nebula flux flares.

\item 4 non-HBLs to provide insights into the blazar sequence, 
to discover their low VHE states, and to study
future VHE flares.
\end{itemize}

\vspace{-0.2cm}
The total targeted exposure on these 14 VHE blazars
is $\sim$220 h per year.  A discovery program
will also continue, with a goal of $\sim$100 h per year.
It will focus on Fermi-LAT selected
objects but will have a larger emphasis on higher-risk / higher reward targets
that may require deep exposures.  An active ToO program on both
known VHE sources and VHE discovery candidates will also be maintained.
The triggers will be based on optical, X-ray, MeV-GeV gamma-ray and
TeV gamma-ray flares.  The VERITAS collaboration hopes to 
increase the amount of blazar ToO data to $\sim$100 h per year.
 
\vspace{0.2cm}
\noindent{\footnotesize  {\bf Acknowledgments}  This research is supported by grants from the U.S. Department of Energy Office of Science, the U.S. National Science Foundation and the Smithsonian Institution, by NSERC in Canada, by Science Foundation Ireland (SFI 10/RFP/AST2748) and by STFC in the U.K. We acknowledge the excellent work of the technical support staff at the Fred Lawrence Whipple Observatory and at the collaborating institutions in the construction and operation of the instrument.}


\end{document}